\documentclass[amsfonts,aps,preprint,nofootinbib]{revtex4-1}
\usepackage{hyperref}
\usepackage[latin1]{inputenc}
\usepackage{graphicx,color}
\usepackage{latexsym}
\usepackage{amsmath}
\usepackage{amssymb}
\usepackage{mathbbol}
\usepackage{fancyhdr}
\usepackage{color}
\newcommand\ee{\end{eqnarray}}
\newcommand\be{\begin{eqnarray}}
\newcommand\ba{\begin{array}}
\newcommand\ea{\end{array}}
\newcommand\beq{\begin{equation}}
\newcommand\eeq{\end{equation}}

\begin{document}
\title{Entanglement from Dissipation and Holographic Interpretation}
\author{M. Botta Cantcheff}
\email{bottac@cern.ch, botta@fisica.unlp.edu.ar}
\affiliation{IFLP-CONICET CC 67, 1900,  La Plata, Buenos Aires, Argentina}
\author{Alexandre L. Gadelha}
\email{agadelha@ufba.br}
\affiliation{Instituto de F\'{\i}sica, Universidade Federal da Bahia,
Campus Universit\'ario de Ondina, CEP: 40210-340, Salvador, BA, Brasil}
\author{D\'afni F. Z. Marchioro}
\email{dafni.marchioro@unila.edu.br , dafnifernanda@gmail.com}
\author{Daniel Luiz Nedel}
\email{daniel.nedel@unila.edu.br}
\affiliation{Universidade Federal da Integra\c{c}\~{a}o Latino-Americana, Instituto Latino-Americano de Ci\^{e}ncias da Vida e da Natureza, Av. Tancredo Neves 6731 bloco 06, CEP: 85867-970, Foz do Igua\c{c}u, PR, Brasil}

\begin{abstract}
In this work we study a dissipative field theory where the dissipation process is manifestly related to dynamical entanglement and put it in the holographic context. Such endeavour is realized by further development of a canonical approach to study quantum dissipation, which consists of doubling the degrees of freedom of the original system by defining an auxiliary one. A time dependent entanglement entropy for the vacumm state is calculated and a geometrical interpretation of the auxiliary system and the entropy  is given in the context of the AdS/CFT correspondence using the Ryu-Takayanagi formula. We show that  the dissipative dynamics is controlled by the entanglement entropy and there are two distinct stages: in the early times the holographic interpretation requires some deviation from classical General Relativity; in the later times the quantum system is
 described as a wormhole, a solution of the Einstein's equations near to a maximally extended black hole with two asymptotically AdS boundaries. We focus our holographic analysis in this regime, and suggest a mechanism similar to teleportation protocol to exchange (quantum) information between the two CFTs on the boundaries (see \cite{maldanovo}).
\end{abstract}

\maketitle

\section{Introduction}
The AdS/CFT correspondence is the most successful application of the holographic principle, and it plays a very important role in the study of the non perturbative sector of a class of Yang-Mills theories. It allows to calculate non-perturbative vacuum expectation values of Yang-Mills operators via tree level calculation in the perturbative sector of type IIB string theory on Anti-de Sitter background. In fact, the conjecture enables calculations of Yang-Mills expectation values in terms of classical propagators on asymptotically AdS geometries. On the other hand, since the AdS/CFT correspondence is a duality, it would be possible to read the holographic dictionary in another way and use the Yang-Mills computations to infer classical properties of the bulk geometry. A more challenging application is to use the holographic dictionary to analyze quantum aspects of gravity. However, our understanding on the basic principle of the holography, even in the context of string theory, is still limited. It is well-known that, to address fundamental issues in quantum gravity such as black hole singularity or quantum cosmology, it is necessary to develop a more general framework for the holography, which allows us to go beyond the Anti-de Sitter spacetime and explore non-trivial generalizations of the AdS/CFT correspondence.

A key property of holography is that it allows to relate geometric quantities
with microscopic information of the quantum field theory. The most famous
example is the Bekenstein-Hawking entropy, which relates the area of a black
hole's event horizon to its entropy: the entropy is directly computable from the
quantum state of the system. In an outstanding application of the AdS/CFT
dictionary,  Ryu and Takayanagi (RT) proposed in \cite{RT} that the area of a
minimal surface in a holographic geometry should be related to the entanglement between the degrees of freedom contained within this region and those of its complement. In particular, they proposed the formula
\begin{equation}
S= \frac{A}{4G_N},
\label{rtformula}
\end{equation}
where S is the entropy of a spatial region $\Omega$ and $A$ is the area of the minimal surface in the bulk whose boundary is given by $\partial \Omega$. Their proposal has been checked in many ways and this relation is proved in \cite{Casini} for a spherical symmetry case and for a more general case in \cite{Lewkowycz}. Quantum corrections to this area law formula are considered in \cite{barela, faulk} and the time evolution of the entropy was studied in \cite{covariantEntropy, covariant}. Also, a holographic entanglement entropy in a higher derivative gravity theory has been obtained in several works, in
particular in \cite{Fusaev}, \cite{Hung, parna, dong, camps, miao}.

The main goal of the present work is to explore the RT formula in the context of dissipative quantum field theories. Dissipative quantum field theories have been studied in the last years for several reasons, and one of these reasons is the experimental results that came out of the Relativistic Heavy Ion Collider (RHIC). The experimental results suggest that the quark gluon plasma is ``strongly coupled" with a very small value of $\eta/s$, where $\eta$ is the shear viscosity and $s$ the thermodynamical entropy density. This result motivates the development of finite temperature holographic techniques to calculate transport coefficients, and a universal result for the shear
viscosity, $\eta/s= 1/(4\pi)$, was derived in \cite{viscosity}. On the other hand,
using the Ba\~nados-Teitelboim-Zanelli (BTZ) black hole \cite{btz} as its
holographic dual, the Brownian motion of a heavy quark in a finite temperature Conformal Field Theory (CFT) was studied in \cite{barnege} in the context of Langevin equation. The holographic Schwinger - Keldysh method was developed for this case in \cite{boer} and the Thermo Field Dynamics (TFD) in \cite{nos}. One of the interesting results is that there is a drag force on the fluctuating external quark even at zero temperature, owing to radiation. This was studied in detail in \cite{barnege2}, where the same result for the zero temperature term is obtained for pure AdS in all dimensions. This suggests that, even at zero temperature, important dissipative effects can be read
from holographic techniques.

A close relation between dissipation and entanglement was pointed out in
\cite{EntanglementDissipation}, where it was reported an experiment where
dissipation generates continuously entanglement between two macroscopic objects. Concerning the quantum treatment of dissipative systems, the quantum theory of damped linear harmonic oscillators can lead to a vacuum state that is in fact an entangled state \cite{fechbacr}. By putting forward the Feshbach-Ticochinski approach in \cite{vitiello}, it was realized a relationship between the canonical quantization of the damped harmonic oscillator and the TFD formalism, where the damped harmonic oscillator was interpreted as the thermal vacuum and the TFD entropy operator appears naturally as the entanglement entropy operator, in a suitable extension to quantum fields. Following these ideas, we are going to use in this work the RT formula to give a holographic interpretation of  the relationship between entanglement and dissipation, concerning conformal field theories at zero and finite temperature.

In order to handle dissipation using a canonical approach, one needs settings
which are wider than those of usual applications of quantum field theory and the AdS/CFT correspondence. In particular, if one introduces dissipation from the outset in the usual settings, two immediate problems come to light. The first one is the  Lorentz covariance breaking: dissipation is a typical process that breaks the invariance under boost transformations. Just because there is a deep relation between dissipation and the ``arrow of time", a dissipative process defines a natural preferred frame. Although it is possible to write a covariant action, the expectation values break the Lorentz covariance \cite{parentani, adamek, Kiritsis}. Also, as thermal effects are taken into account, the thermal bath's frame works as the preferred frame. The second problem is the loss of unitarity. However, this problem can be circumvented and it is possible to preserve the unitarity and make usual quantum mechanics in a finite volume limit. In order to do such endeavour, we shall work therefore with Hamiltonian theories in which dissipative effects are caused by interactions with additional degrees of freedom \cite{vitiello, parentani, adamek}. In this case, the AdS/CFT correspondence allows an elegant interpretation of the auxiliary degrees of freedom. The explanation follows the same spirit of Israel's work in black holes using TFD \cite{Israel}. In TFD, in order to take care of thermal expectation values, the original system is duplicated and the thermal vacuum is defined via a Bogoliubov transformation, which actually entangles the system and its copy. In maximally extended black hole solutions, the auxiliary degrees of freedom are interpreted as fields living behind the horizon; for AdS black holes, this defines two asymptotic conformal theories. In references \cite{TraversableWormholes, maldanovo}, a mechanism  to send a signal between the boundaries is presented. From a teleportation protocol, an  effective potential reproduces an ``attractive force" that brings the boundaries  closer . The potencial is set up with the product of two hermitian operators, one on each side.  In our work, we obtain a similar potential that couples the two boundaries.

We consider a particular coupling between two systems that will drive dissipation in one of the boundaries. Note that what defines the system and the auxiliary system is just the geometric region where the measures are taken. We show that, for one class of asymptotic observers, the vacuum state is an entangled state and we calculate the entanglement entropy in a ``canonical way". Actually, we show that the entanglement entropy can be calculated via expectation value of an operator, defined as the entanglement entropy operator, which is in fact a representation for the modular Hamiltonian \cite{haag}, playing a dynamical role in this system. In fact, it will be seen that the time evolution of the vacuum entangled state is generated by the time derivative of that operator. So, after establishing a geometric description for the dissipative conformal field theory (DCFT), we can use the RT formula to infer that the time evolution of the minimal surface is controlled by the time dependence of the entangled state.

We observe at least two distinct holographic stages: for the first one, at very early times, some type of deviation from classical Einstein's gravity should be considered in order to describe changes of the spacetime topology \cite{collapse}; in the second one, for later times, we can study the leading large-N effects by considering a classical solution of the Einstein's equations as the holographic dual. The final sections of the present paper will be devoted to study the second holographic stage.

In order to find out the geometric picture of the dissipative scenario presented  here, we look for an asymptotically AdS gravitational system that reproduces the same dissipative dynamics on the boundary.  It is known for a long time that scalar field coupled to Friedmann-Robertson-Walker (FRW) metric  has a kind of dissipative behavior \cite{dissFRW, grazi1, grazi2, buryak, habib}. In particular, the equation of motion has the same damping term of the DCFT studied here. In \cite{Tetradis} it was shown that, using a particular coordinate transformation, it is possible to have FRW metric on the boundary from a BTZ black hole on the bulk.  So, this appears to be the geometric system that we are looking for. However, it will be shown that the time dependent entanglement entropy  derived here demands that the original metric is not BTZ, but a sort of Vaidya solution in the adiabatic approximation; that is, a BTZ black hole with a slowly time dependent mass. Keeping this scenario in mind, the RT formula allows to find a natural relationship between dissipation, entanglement, thermodynamics and black hole physics. However, as it will be disscussed in section 7, for $AdS_3$ the  holographic picture based on the FRW bonudary is just an approximation of the DCFT. In the asymptotic limit, the entangled state derived here can be seen as an approximation of the TFD vacuum.

This work is organized as follows: in section 2 we present the dissipative model; in sections 3 and 4 the time dependent entropy is canonically computed; in section 5 we show how we deal with the time dependence of the entropy in holographic computations; in section 6 the holographic model is constructed; and section 7 is devoted to the conclusions.

\section{The Dissipative Model}

In this section we are going to develop the main idea of this work. We start with a system A, whose degrees of freedom are described by a free conformal field $\phi$ living on a compact space $\Omega \subset \Re^d$. The usual canonical approach to study dissipative systems consists of putting the system $A$ inside another system $\bar{A}$ (bath, medium, or environment), whose degrees of freedom are unknown. The system $\bar{A}$ interacts with the system $A$ such that, from the $\phi$-field perspective, there is a dissipative process which can be described by the equation
\begin{equation}
(\partial_t^2-\nabla^2 )\phi+ \gamma\partial_t\phi=0,
\label{osci1}
\end{equation}
where $\gamma$ is the damping coefficient that will describe the coupling between the original  fields and the medium (the $\bar{A}$ system). Here we study the problem from a different perspective. We think of A and $\bar{A}$ as two physical systems living in different geometric regions, corresponding to two asymptotic conformal theories. In a certain moment an interaction is suddenly switched on, allowing energy exchange between the two theories\footnote{This may be viewed as a sort of quantum quenching
\cite{myers}, where an interaction is suddenly turned on.}. The equation (\ref{osci1}) is just the resulting dynamics as seen by the A system.

If we represent the $\bar A$ system by $\psi$ fields, we can write the following Lagrangian that reproduces the equation (\ref{osci1})

\begin{equation}
L= \int_\Omega d^dx\,\psi\,\left[ (\partial_t^2-\nabla^2 )\phi
+ \gamma\partial_t\phi\right].
\label{lagrange-psi}
\end{equation}

\noindent Integrating (\ref{lagrange-psi}) by parts and neglecting the boundary terms, we can write it in the symmetric form

\begin{equation}
L= \int_\Omega d^dx \left[\partial_{\mu}\phi\partial^{\mu}\psi
+ \frac{\gamma}{2}\left(\phi\partial_t\psi-\psi\partial_t\phi\right)\right].
\label{Lagrangeano}
\end{equation}

\noindent The variation of (\ref{Lagrangeano}) with respect to $\psi$ gives (\ref{osci1}), while minimizing it with respect to $\phi$ leads to the equation of motion for the  $\psi$ field
\begin{equation}
(\partial_t^2-\nabla^2 )\psi- \gamma\partial_t\psi=0.
\label {osci2}
\end{equation}
Equation (\ref{osci2}) shows that $\psi$ describes an identical copy of the physical field $\phi$ with the time direction reversed. The interaction between $\psi$ and $\phi$ describes the dissipative behavior precisely \cite{vitiello, parentani, adamek}.

We want to draw attention for the similarity of this construction with the rules
of TFD: the field $\psi$ can be considered the TFD's double of the system, and it can be interpreted as a derivation of the TFD picture. In this context, the field $\psi$ evolves in the inverse time direction. In the usual interpretation, if the coupling is switched off ($\gamma\to 0$), we recover two non-interacting (free) CFT's whose states describe spacetimes with two locally asymptotically AdS regions \cite{collapse}. In particular, the ground state $|0>_\phi \otimes |0>_\psi$ represents two disconnected copies of the exact AdS spacetime \cite{VR}, and the thermal vacuum\footnote{In a TFD description.} is the  Kruskal extension of an eternal AdS black hole\footnote{For further applications of TFD in string theory, see \cite{nos} and \cite{IonRev}.} \cite{eternal}. Furthermore, the AdS/CFT dictionary implies that, in this example, the gravity dual theory is very stringy, and therefore quantum gravity interprets it as quantizing strings on AdS such that these metrics shall be recovered in the proper large N-limit. In the following sections, the behavior of the entropy will refer us to two different scenarios. We will focus on the large $\gamma t$ limit, where it will be possible to make an approximation compatible with the large N limit, and the holographic picture can be understood in terms of Vaidya black holes, written in a particular coordinate system. Also note that the Lagrangian (\ref{Lagrangeano}) is different from the ones studied in \cite{TakacoupledCFTs, mozafar} just because we are interested in the dissipative process and its relation with the entanglement entropy. As the dissipative process imposes a privileged direction of time, Lorentz invariance is broken \cite {parentani, adamek, Kiritsis}. However, the entanglement entropy for models with broken Lorentz symmetry has the same behavior as the relativistic ones \cite{Solodukhin}.

The canonical conjugate momenta of the fields are
\begin{eqnarray}
\pi_{\phi}&=& \frac{\partial L}{\partial\dot\phi}=
\dot\psi- \frac{\gamma}{2}\psi
\nonumber
\\
\pi_{\psi}&=& \frac{\partial L}{\partial\dot\psi}= \dot\phi+ \frac{\gamma}{2}\phi,
\end{eqnarray}
and the Hamiltonian can be written as
\begin{equation}
\mathcal{H} = \pi_{\phi}\pi_{\psi}
+\partial_i\phi\partial^i\psi-\frac{1}{2}\gamma(\phi\pi_{\phi}-\psi\pi_{\psi})
-\frac{\gamma^2}{4}\psi\phi.
\end{equation}
The ansatz for the solution of the equations of motion is
\begin{equation}
\phi= \phi' e^{-\frac{\gamma}{2}t},
\end{equation}
where $\phi'$ is the plane wave solution with frequency
$\omega_{k} = \pm \,\sqrt{k^2  - \frac{\gamma^2}{4}}$. Note that the system has real frequencies since an IR cut off is defined by $\gamma$.

The general solution can be expanded in terms of quasi-normal frequencies
(frequencies with imaginary part), namely
\begin{equation}
\phi= \phi_\Omega(x) e^{i\Omega t}, \qquad \Omega = w + i \Gamma.
\end{equation}
Replacing this into the equation of motion and demanding that $k$ is real\footnote{Otherwise we would have that $\Gamma$ depends on $k$.}, we have that $ \Gamma = \frac{\gamma}{2}$ and $\omega_{k} = \pm \,\sqrt{k^2  - \frac{\gamma^2}{4}}$. In summary, the general solution is plane waves damped by a decaying factor $e^{-\frac{\gamma}{2}t}$. In contrast, the solution for the field $\psi$ has a growing factor $e^{\frac{\gamma}{2}t}$; however,  by taking the physical time parameter $-t$, it has the correct damping behavior.

It is easier to quantize the theory by making the redefinition
\begin{equation}
\Phi = \frac{\phi+\psi}{\sqrt{2}},\qquad \Psi=\frac{\phi-\psi}{\sqrt{2}}.
\end{equation}
In terms of the new fields the Lagrangian is
\begin{equation}
L = \int d^dx \left\{\frac{1}{2} \left[\left(\partial_{\mu}\Phi\right)^2
- \left(\partial_{\mu}\Psi\right)^2 \right]
+ \frac{\gamma}{2} \left(\Psi \dot{\Phi} - \Phi \dot{\Psi}\right)\right\},
\end{equation}
and the canonical momenta are
\begin{equation}
\Pi_{\Phi}= \dot\Phi + \frac{\gamma}{2}\Psi, \qquad \Pi_{\Psi}
= -\dot\Psi- \frac{\gamma}{2}\Phi.
\end{equation}
Now the Hamiltonian can be represented as
\begin{equation}
H = H_{\Phi}- H_{\Psi},
\end{equation}
where
\begin{eqnarray}
H_{\Phi} = \frac{1}{2} \int d^{d}x \left[\left(\Pi_{\Phi}
- \frac{\gamma}{2} \Psi\right)^2 + \frac{1}{2} (\partial_i \Phi)^2\right]
\nonumber
\\
H_{\Psi} = \frac{1}{2} \int d^{d}x \left[\left(\Pi_{\Psi}
+ \frac{\gamma}{2} \Phi\right)^2  + \frac{1}{2} (\partial_i \Psi)^2 \right].
\label{hcorrec}
\end{eqnarray}

In momentum space, the Hamiltonian is
\begin{equation}
H= \int d^dk \left[\bar{\Pi}_{\Phi}(k) \bar{\Pi}_{\Phi}(-k) -\bar{\Pi}_{\Psi}(k) \bar{\Pi}_{\Psi}(-k) +\frac{1}{2}k^2(\Phi(k)\Phi(-k) -\Psi(k)\Psi(-k))\right],
\end{equation}
where
\begin{equation}
\bar{\Pi}_{\Phi} = \Pi_{\Phi} - \frac{\gamma}{2} \Psi \qquad
\bar{\Pi}_{\Psi} =  \Pi_{\Psi} + \frac{\gamma}{2} \Phi .\nonumber
\end{equation}

As the total system is canonical, the fields satisfy the usual canonical
commutation relations and they can be replaced by expressions in terms of the creation/annihilation operators. As usual, the free fields can be written as
\begin{eqnarray}
\Phi (k) = \frac{A_k + A^{\dagger}_{-k}}{\sqrt{2\epsilon_k}} ,
\qquad \Pi_{\Phi} (k) = \frac{\sqrt{2\epsilon_k} (A_k - A^{\dagger}_{-k})}{2i},
\nonumber
\\
\Psi (k) = \frac{B_k + B^{\dagger}_{-k}}{\sqrt{2\epsilon_k}} ,
\qquad \Pi_{\Psi} (k) = \frac{\sqrt{2\epsilon_k} (B_k - B^{\dagger}_{-k})}{2i},
\end{eqnarray}
where $[A(k), A^{\dagger}(p)] = [B(k), B^{\dagger}(p)] = \delta(k-p)$ and  $\epsilon_k = \sqrt{k^2}$. The momentum space Hamiltonian becomes
\begin{eqnarray}
H = \sum_k  \left[\left(\epsilon_k-\frac{\gamma^2}{8\epsilon_k}\right)
\left(A^{\dagger}_k A_k - B^{\dagger}_k B_k\right)- \frac{\gamma}{2i}
\left(A_{-k} B_{k} - A^{\dagger}_{-k} B^{\dagger}_k\right)
\right.
\nonumber
\\
\left.
+ \frac{\gamma^2}{16\epsilon_k}\left(B_{k}B_{-k}
+ B^{\dagger}_{k}B^{\dagger}_{-k}- A_{k}A_{-k}
- A^{\dagger}_k A^{\dagger}_{-k}\right) \right].
\label{han}
\end{eqnarray}
It is easy to see that the Hamiltonian in (\ref{han}) can be presented as
\begin{equation}
H = H_{0} + H_{int},
\end{equation}
with
\begin{eqnarray}
H_{0} &=& \sum_{k}
\left(\epsilon_k-\frac{\gamma^2}{8\epsilon_k}\right)
\left(A^{\dagger}_{k} A_{k} - B^{\dagger}_{k} B_{k} \right),
\nonumber
\\
H_{int} &=& \sum_{k} \left[-\frac{\gamma}{2i} \left(A_{-k} B_{k}
- A^{\dagger}_{-k} B^{\dagger}_{k}\right)
+\frac{\gamma^2}{16 \epsilon_k}\left(B_{k}B_{-k}+
B^{\dagger}_{k}B^{\dagger}_{-k} - A_{k}A_{-k}
- A^{\dagger}_{k}A^{\dagger}_{-k} \right)\right].
\nonumber
\end{eqnarray}
As $\left[H_{0},H_{int}\right] = 0$, the vacuum state time evolution is given by \footnote{Actually, $\left|0 \right\rangle$ is a $SU(1,1)$ rotated vacuum. See \cite{vitiello,vitiello2, fechbacr, solomon} for details.}
\begin{eqnarray}
\left|0(t) \right\rangle &=& \exp(-i H t) \left|0 \right\rangle
\nonumber
\\
&=& \exp\left[\sum_{k}\; \frac{\gamma t}{2} \left(A_{-k} B_{k}
- A^{\dagger}_{-k} B^{\dagger}_k\right)\right]
\left|0 \right\rangle
\nonumber
\\
&=& \prod_{k} \frac{\delta_{kk}}{\cosh\left(\frac{\gamma t}{2}\right)}
\exp\left[\tanh\left(\frac{\gamma t}{2}\right)
A^{\dagger}_{-k}B^{\dagger}_{k}\right]\left| 0 \right\rangle.
\label{res1}
\end{eqnarray}
Note that the dissipative interaction of the two CFTs produces an entangled state. In the next section we are going to show how the entanglement entropy appears naturally in this scenario. Also note that this state becomes stable for $t \gg 4 / \gamma$ \cite{malda-hartman},
\begin{eqnarray}
\left|0(t) \right\rangle \sim N \exp\left[ A^{\dagger}_{-k}B^{\dagger}_{k}\right]\left| 0 \right\rangle \equiv \left|B \right\rangle,
\label{finalBstate}
 \label{eas}
\end{eqnarray}
where $\left|B \right\rangle$ is a boundary state in the field theory.  In
particular this is the solution for the equation
\begin{equation}
\left[\Phi(x,t) - \Psi(x,t)\right ] \left|B \right\rangle=0.
\end{equation}

In terms of the original fields, this translates into $\phi \left|B \right\rangle= \psi \left|B \right\rangle=0$, as we should expect in a dissipative theory if we do not take into account thermal effects;  this final state is the maximum entanglement state for each mode. In fact, the classical behavior of the fields for later times translates into a (quantum) boundary condition that reads
\begin{eqnarray}
\phi \left| 0(t) \right\rangle \approx 0,
\label{BCfisico}
\end{eqnarray}
for $ 4 / \gamma \ll t $, while the state $\left|B \right\rangle$ is actually defined by the exact condition
\begin{eqnarray}
\phi \left| B \right\rangle = 0.
\label{BC}
\end{eqnarray}
However, thermal effects are always present in dissipative theories \cite{livro}. The Brownian motion does not allow a condition like (\ref{BCfisico}) and (\ref{eas}). This implies that, at least in later times, the $\gamma$ must not be independent of modes and the asymptotic limit for the vacuum state cannot be (\ref{eas}).

\section{Entanglement Entropy and Dissipative Dynamics}

In order to analyze the result of equation (\ref{res1}) in a more general
context, let us consider for a moment a possible dependence of the dissipative coefficient on the particles' momenta, namely $\gamma \to \Gamma_k$. Following reference \cite{vitiello}, the entanglement entropy can arise in the scenario introduced above observing that the vacuum at a finite time $t$, in a finite volume, can also be obtained as
\begin{eqnarray}
\left|0(t) \right\rangle
&=& e^{-\frac{1}{2} S_{A}(t)} e^{\sum_{k} A^{\dagger}_{-k}B^{\dagger}_{k}}
\left| 0 \right\rangle
\\
&=& e^{-\frac{1}{2} S_{B}(t)} e^{\sum_{k} A^{\dagger}_{-k}B^{\dagger}_{k}}
\left| 0 \right\rangle ,
\label{estadoentropia}
\label{exp-entropia}
\end{eqnarray}
where we have introduced the following operators
\begin{eqnarray}
S_{A}(t) = - \sum_{k} \left[ A^{\dagger}_{k}A_{k}\ln\sinh^{2}\left(\Gamma_{k} t\right)
- A_{k}A^{\dagger}_{k}\ln\cosh^{2}\left(\Gamma_{k} t\right)\right],
\\
S_{B}(t) = - \sum_{k} \left[B^{\dagger}_{k}B_{k}\ln\sinh^{2}\left(\Gamma_{k} t\right)
- B_{k}B^{\dagger}_{k}\ln\cosh^{2}\left(\Gamma_{k} t\right)\right].
\label{entAB}
\end{eqnarray}
Many interesting properties are related to these operators.\footnote{These operators were first studied in TFD context in \cite{chuume} and further developed in \cite{gadened1}. The relation with dissipative dynamics was first studied in \cite{vitiello}.} First of all, considering the result of their expectation value in the time evolved vacuum,
\begin{equation}
\left\langle 0 \left(t\right)\left| A^{\dagger}_{k}A_{k} \right|0 \left(t\right) \right\rangle
= \left\langle\ 0 \left(t\right)\left| B^{\dagger}_{k}B_{k} \right|0 \left(t\right) \right\rangle
= \sinh^{2}\left(\Gamma_{k}t\right),
\end{equation}
we can write it as
\begin{equation}
{\cal S}_{A}(t)= \left\langle 0 \left(t\right)\left| S_{A} \right|0 \left(t\right) \right\rangle = - \sum_{n\geq 0}{\cal W}_{n}(t) \ln {\cal W}_{n}(t),
\label{entA-t}
\end{equation}
where
\begin{equation}
{\cal W}_{n}(t)=\prod_{k}
\left(\frac{\sinh^{2n_{k}} \left(\Gamma_{k}t\right)}{\cosh^{2n_{k}+2} \left(\Gamma_{k}t\right)}\right),
\end{equation}
and the same can be done for $S_{B}$. Notice also that the vacuum at a finite $t$ can be written in terms of ${\cal W}_{n}(t)$

\begin{equation}
\left|0(t) \right\rangle= \sum_{n} \sqrt{{\cal W}_{n}(t)}\left|n,n \right\rangle.
\end{equation}

The expressions above allow us to make an important connection with the density operator. Considering
\begin{equation}
\rho = \left|0(t) \right\rangle \left\langle 0(t)\right|.
\end{equation}
then the reduced density operator is
\begin{eqnarray}
\rho_{A} &=& Tr_{B}\left[ \left|0(t) \right\rangle \left\langle 0(t)\right| \right]
\nonumber
\\
&=& \sum_{n} {\cal W}_{n}(t)\left| n \right\rangle \left\langle n \right|
\end{eqnarray}
that is, the entropy operator coincides with the entanglement Hamiltonian defined in the literature \cite{haag}. In fact, by tracing out the $B$-degrees of freedom, we obtain the time-dependent reduced matrix density \cite{area-qg}
\begin{equation}
\rho_A (t) = e^{-{\cal S}_A (t)},
\end{equation}
showing that the entropy operator ${\cal S}_A (t)$ is nothing but the so-called modular Hamiltonian for this state \cite{haag}. Then, the expectation value of any $A$-operator ${\cal O}_A$ in $\left|0(t) \right\rangle$ can be written as
\begin{equation}
\left\langle 0(t)| {\cal O}_A |0(t) \right\rangle= Tr \rho_A {\cal O}_A
\end{equation}

Using equation (\ref{estadoentropia}), we can verify that the time evolution of the vacuum state is generated by the time derivative of the entropy operator or entropy production, namely
\begin{equation}
\frac{\partial \left| 0(t) \right\rangle}{\partial t}= -\frac{1}{2}\frac{\partial S}{\partial t}  \left| 0(t) \right\rangle,
\label{eq-mov-entropy}
\end{equation}
which has an interesting holographic interpretation, as it will be discussed in the next section. Remarkably, let us observe that this equation resembles thermodynamics, even though its derivation has nothing to do with the thermodynamic laws.  It is interesting how, by virtue of this equation, the basic notion of equilibrium $\displaystyle\frac{\partial \left| 0(t) \right\rangle}{\partial t} \approx 0$ is equivalent to the maximum entropy condition
\begin{equation}
\frac{\partial S}{\partial t}  \left| 0(t) \right\rangle \approx 0  ,
\end{equation}
which is fulfilled if the operator $S$ achieves its maximum value for some time $t$. This might be recognized precisely as the second law of thermodynamics in a quantum-mechanical sense. For instance, in the example given previously, we can observe explicitly that
\begin{equation}
\frac{\partial S}{\partial t} \approx 0 \qquad \mathrm{as}
\qquad t \gg 4 / \gamma .
\end{equation}
This can be easily seen from equation (\ref{exp-entropia}) where
$e^{-\frac{1}{2} S_{A}(t)} \approx I$ for $t \gg 4 / \gamma$. We shall return to this point in section V.\footnote{In \cite{termorelation}  it is argued that the entanglement entropy for a very small subsystem obeys a property which is analogous to the first law of thermodynamics; here this kind of relation naturally arises.}

\subsection{DCFT renormalized entanglement entropy}

Let us return now to our original example, where $\Gamma_k \equiv \gamma\,\,\, \forall\,\, k$. In this case, the expressions above simplify considerably and we can compute explicitly the entanglement entropy as a function of time. In fact,
\begin{eqnarray}
S_{A}(t) &=& -
\sum_{k} \left[ A^{\dagger}_{k}A_{k}\ln\sinh^{2}\left(\gamma t\right) - A_{k}A^{\dagger}_{k}\ln\cosh^{2}\left(\gamma t\right)\right]
\nonumber
\\
&=& -\hat{N}_{A} \left[\ln\sinh^{2}\left(\gamma t\right)
- \ln\cosh^{2}\left(\gamma t \right)\right]
+ \sum_{k}\ln\cosh^{2}\left(\gamma t\right)
\label{entrconta1},
\end{eqnarray}
where
\begin{eqnarray}
\hat{N}_A = \sum_{k}  A^{\dagger}_{k}A_{k}.
\end{eqnarray}
Defining the functions
\begin{eqnarray}
F(t) := - \ln \tanh^{2}\left(\gamma t\right),\\
G(t) := \sum_{k}\ln\cosh^{2}\left(\gamma t\right),
\end{eqnarray}
then the entanglement entropy operator reads simply as
\begin{eqnarray}
S_{A}(t) =  F(t) \hat{N}_A + G(t) .
\label{entrconta-f}
\end{eqnarray}
The equation (\ref{entrconta-f}) is very useful since, in principle, we
can follow the change in the area of an extremal surface in a dual holographic geometry. Actually, this is an operator whose expectation value in the time dependent state defined in (\ref{res1}) provides the area of an extremal surface. For instance, taking the expectation value in the time dependent ground state, we have
\begin{eqnarray}
s_{A}(t) &=& \sum_{k} \delta_{kk} [- \sinh^{2}\left(\gamma t\right)\, \ln \tanh^{2}\left(\gamma t\right)
-\ln\cosh^{2}\left(\gamma t\right)] \nonumber\\
\label{entrcontaxxxx}
\end{eqnarray}
Note that when the volume of the space is finite, there is an UV finite term in entropy densities. Differently from the UV behavior of the usual entanglement entropy (which scales with the area), one expects the leading UV behavior of this entanglement entropy to scale with the volume \cite{Marika}. In order to obtain a finite result, we introduce an  UV regulator. Using periodic  boundary conditions in $d=2$, the result is
\begin{eqnarray}
{\cal S}(t) = -2\sum_{n=1}^{\infty}e^{-\epsilon 2\pi n/L}
\left[\sinh^2\left(\gamma t\right)\ln\tanh^{2}\left(\gamma t\right)
- \ln\cosh^2\left(\gamma \right )\right]
\nonumber
\\
= -\left[ \sinh^{2}\left(\gamma t\right) \,\ln\tanh^{2}\left(\gamma t\right)
- \ln\cosh^2\left(\gamma t\right)\right]
\left(\frac{L}{\pi\epsilon} - 1 + O(\epsilon/L)\right),
\end{eqnarray}
where L is the circumference of the $S^1$ space and $\epsilon$ is the UV regulator.  Following the usual procedure, we can pick up the UV finite term by differentiating the entropy density in the limit $\epsilon \rightarrow 0$:
\begin{equation}
\frac{\partial}{\partial L}\left(\frac{{\cal S}(t)}{L}\right) =
-\frac{1}{L^2}\left[\sinh^{2}\left(\gamma t\right)\,\ln\tanh^{2}
\left(\gamma t\right) - \ln\cosh^2\left(\gamma t\right)\right],
\end{equation}
and the renormalized entropy is
\begin{eqnarray}
s^{Ren}_{A}(t) &=&
-\sinh^{2}\left(\gamma t\right)\ln\tanh^{2}\left(\gamma t\right) +\ln\cosh^{2} \left(\gamma t\right)
\nonumber
\\
&=& \mathfrak{f}(t)\sinh^{2}\left(\gamma t\right)+ \mathfrak{g}(t) ,
\label{entrcontaxxxx}
\end{eqnarray}				
where
\begin{equation}
\mathfrak{f}(t)= -\ln\tanh^{2}\left(\gamma t\right), \qquad  \mathfrak{g}(t)= \ln\cosh^2 \left(\gamma t\right) .
\label{fg}
\end{equation}

For $\gamma t \ll 1$ 	
\begin{equation}
s^{Ren}_{A}(t) \approx -(\gamma t)^2\ln \left(\gamma t\right) \label{et}
\end{equation}	
Notice the similarity with the results found in \cite{TakacoupledCFTs, mozafar}.

Now, let us consider the asymptotic limit $\gamma t \gg 1$:
\begin{equation}
s^{Ren}_{A}(t) \approx \ln(\cosh (\gamma t))\approx \gamma t . \label{at}
\end{equation}
In the  asymptotic limit, the entropy grows linearly with time according to the Cardy/Calabrese results. Actually, the asymptotic entropy resembles the finite temperature results of \cite{malda-hartman}, if $\gamma$ is interpreted in terms of the temperature. This suggests that asymptotically the system thermalizes, which is no surprise when dealing with a dissipative system. Before carrying out the holographic interpretation of this result, let us understand how to deal with the time dependence of entropy in holographic computations. This is the main goal of the next section.

\section{The Time Dependent Holographic Computations}

The renormalized entropy found in (\ref{entrcontaxxxx}) would be $4G$ times the area of the minimal surface in the dual holographic space whose boundary is anchored  by $\partial \Omega$. However, one does not know a priori
how much of the variation of this entropy is due to the dual metric change or to
the extremal surface, or both. So, let us discuss this point in detail.
\begin{figure}[!h]
\centering
\includegraphics[scale=0.5]{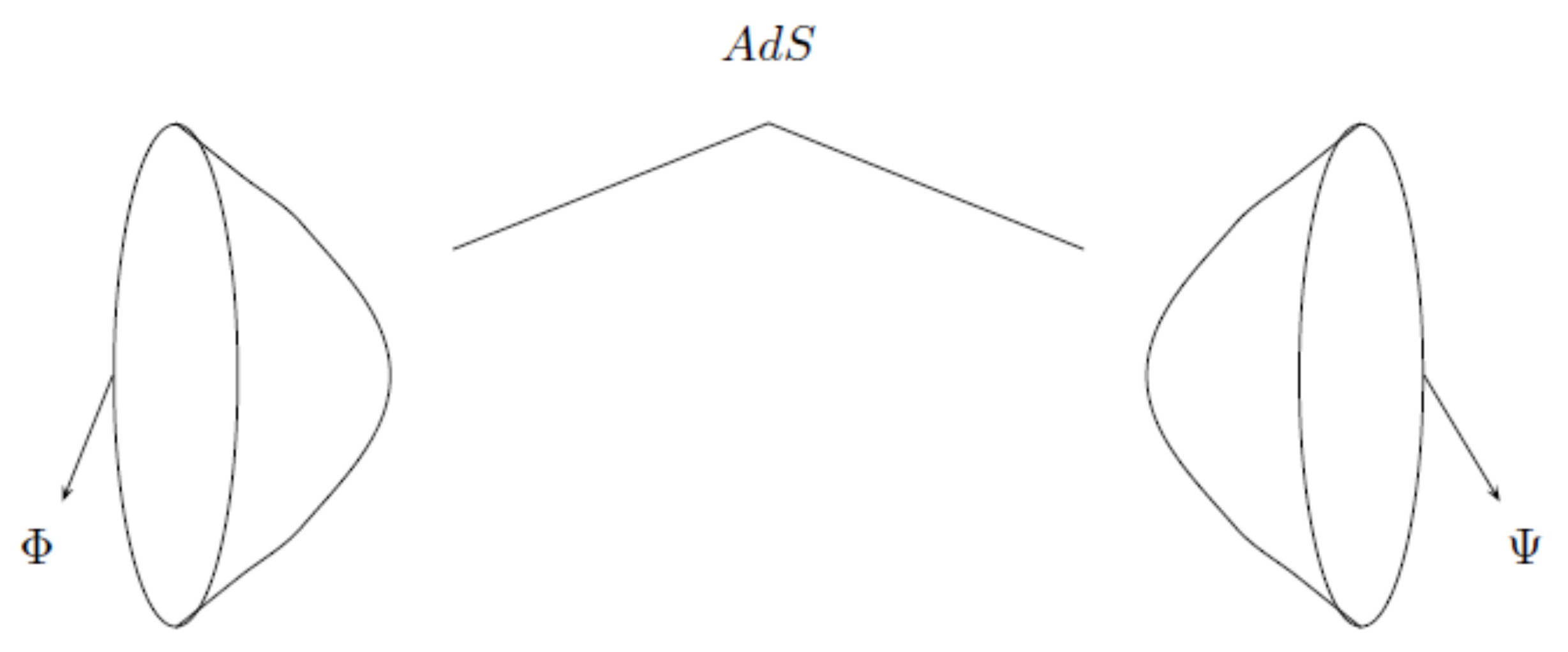}
\caption{The state $\left.\left|0\right\rangle\!\right\rangle
= \left|0\right\rangle_A\otimes\left|0\right\rangle_B$ is dual to two disconnected
copies of AdS.}
\label{fig1}
\end{figure}

According to the AdS/CFT dictionary, the state $\left|0(t)\right\rangle $ is dual
to a background spacetime metric, namely $g_{\mu \nu}(t)$\footnote{We are assuming implicitly that the fields are in the adjoint representation of a SU(N) group in the large N limit.}. In particular, $\left|0(t=0)\right\rangle = \left. \left|0 \right\rangle\!\right\rangle$ corresponds to the globally AdS spacetime; actually, $\left.\left|0\right\rangle\!\right\rangle =
\left |0\right\rangle_A\otimes\left|0\right\rangle_B$ is dual to two disconnected
copies of AdS as in figure \ref{fig1} (see \cite{VR, collapse} for interpretations
of such geometries). On the other hand, the entanglement entropy is given by the RT formula,
\begin{equation}
s(t)= \left\langle 0(t)\right| S_{A}(t) \left|0(t)\right\rangle
= \frac{1}{4G_N} \, a[ g_{\mu\nu}(t) , \Sigma_t ]
:= \frac{1}{4G_N} \int_{\Sigma_t} \sqrt{\det g^{(ind)}(t)},
\label{RT}
\end{equation}
where $g^{(ind)}(t)$ denotes the induced metric on the minimal (co-dimension two) surface $\Sigma_t$. The surface represented by the integral on the right hand side of (\ref{RT}) must be properly regularized with a cut-off such that $\displaystyle\Sigma_t^\epsilon$ has a finite area (see details on this calculation in the context of Wilson loops in \cite{wloops, albash, pontelo}).

The limit case
\begin{equation}
\lim_{ t\to 0}  \left\langle 0(t)\right| S_{A}(t) \left|0(t)\right\rangle = 0,
\label{RT0}
\end{equation}
reflects the fact that the minimal surface $\Sigma_0$, embedded in the exact AdS spacetime (whose conformal boundary is $S^d$) and anchored by $\partial \Omega$, has vanishing area as $\Omega\to S^d$. Then, using (\ref{RT}), the metric induced on $\Sigma_t$ is
\begin{equation}
\mathfrak{f}(t) \sinh^2 \left(\gamma t \right)+ \mathfrak{g}(t) =
\frac{1}{4G_N} \int_{\Sigma_t} \sqrt{\det g^{(ind)}(t)}.
\label{F}
\end{equation}
In other words, a holographic dual of our model consists of a metric
$g_{\mu \nu}(t)$ equipped with a minimal surface $\Sigma_t$ that satisfies
(\ref{F}).

Considering first order variations of the state $\left|0(t)\right\rangle$, we can
verify that
\begin{equation}
\delta s_A (t)
\equiv \delta\left\langle 0(t)\right| S_{A}(t) \left|0(t)\right\rangle  =
- \mbox{Tr}_{A} \, \delta\rho_{A} \, S_{A}(t) + O^2 (\delta \rho_{A})
\label{var1}
\end{equation}
where the variation is normalized as $\mbox{Tr}_A \: \delta \rho_A =  \left\langle 0(t)|\delta 0(t)\right\rangle = 0$. The equation (\ref{var1}) can be written as
\begin{eqnarray}
\delta s_{A}(t)  =
\left\langle 0(t) + \delta 0(t)\left| S_{A}(t)\right| 0(t)
+\delta 0(t) \right\rangle
- \left\langle 0(t)\left| S_{A}(t) \right| 0(t)\right\rangle .
\end{eqnarray}
In particular, one can think of $t$ (or $\gamma t$) as a parameter for the states' metrics and compute
\begin{eqnarray}
\left\langle 0\left| S_{A}(t) \right|0\right\rangle
= \mathfrak{g} (t) = (\gamma t)^2 + O^3(\gamma t) \dots
\end{eqnarray}
For instance,
\begin{eqnarray}
\left\langle 0(t)\left| S_{A}(t) \right|0(t)\right\rangle
- \left\langle 0\left| S_{A}(t) \right|0\right\rangle
= \mathfrak{f}(t) \sinh^2 \left(\gamma t \right),
\end{eqnarray}
while, using (\ref{RT0}),
\begin{eqnarray}
\delta s_{A}(t) \equiv \left\langle 0(t)\left| S_{A}(t) \right|0(t)\right\rangle
- \left\langle 0\left| S_{A}(0) \right|0\right\rangle
= \mathfrak{f}(t) \sinh^2 (\gamma t)  + \mathfrak{g}(t) ,
\end{eqnarray}
and the difference between both expressions is precisely
$O^2 (\delta \rho_A) = (\gamma t)^2 + O^3(\gamma t)$, which expresses
(\ref{var1}) in terms of the parameter $t$ (or $\gamma t$).

On the other hand, since the first variation of the area with respect to
$\Sigma_t$ vanishes \cite{Pando}, we have that
\begin{equation}
\delta s_{A}(t) = Tr_{A} \, \delta \rho_A \, S_{A}(t)
= \frac{1}{4G_N}\left(\frac{\delta a}{\delta g_{jk}}\right)_{\Sigma_t}
\delta g_{jk}(t) := \frac{1}{4G_N} \int_{\Sigma_t}\delta \sqrt{det g^{(ind)}(t)},
\label{delta-2}
\end{equation}
where, since the co-dimension $\Sigma_t$ is two, the area is entirely embedded in a spacelike hypersurface ${\cal N}$, so here $g_{jk}$ stands for the spacelike Riemannian metric on ${\cal N}$.

Therefore, one can conclude that for states
$\left|\psi\right\rangle \equiv \left|0(t) +  \delta 0(t)\right\rangle$ near to
$\left|0(t)\right\rangle$ the interpretation of the quantity
\begin{equation}
\left\langle \psi\left| S_{A}(t) \right|\psi\right\rangle
= \frac{1}{4G_N} \, a[ g_{jk}(t) + \delta g_{jk}, \Sigma_t ]
:= \frac{1}{4G_N} \int_{\Sigma_t} \sqrt{det (g + \delta g)^{(ind)}(t)}
\end{equation}
is the area of the same surface $\Sigma_t$ calculated in the deformed metric $g^{(\psi)}\equiv g_{jk}(t) + \delta g_{jk}$. Note that $\Sigma_t$ is the extremal one for the metric $g_{jk}(t)$, then in general it doesn't need to be the extremal surface for $g^{(\psi)}$. Actually, for states arbitrarily different from $\left|0(t)\right\rangle$, one does not know on which surface should evaluate this functional, nor if the area law is even valid or there is corrections to it. Nevertheless, one can find a holographic (differential) formula that may be a useful recipe to track the dual metric in certain
specific cases, where the minimal embedded surface does not change as the metric is deformed. In principle one must consider a one-parameter family of Riemannian metrics $\{g_{jk}(\theta)\}_{\theta\in R}$ on a fixed manifold (space) ${\cal N}$, but in the cases we are going to study, we can identify the real parameter $\theta$ with $t$, or $ \gamma t$.

We can differentiate the expression (\ref{delta-2}) with respect to the time
parameter and obtain the holographic formula for the entropy production
\begin{equation}
\sigma^R \equiv  \frac{d s(t)}{d t}
= \frac{1}{4G_N} \, \left( \frac{\delta a}{\delta g_{jk}}\right)_{\Sigma_t }
\, \frac{d g_{jk}}{d t},
\label{holographic-sigma}
\end{equation}
where the time/parameter derivative is the total derivative since $t$ shall be interpreted as a parameter rather than a spacetime coordinate. However, if the whole spacetime is built up as a foliation of spaces
$\{({\cal N} , g_{jk})_t, \}_t \equiv M$, there is a frame where the parameter $t$
is the time coordinate and it is on the same foot as the other coordinates of the
spacetime,
\begin{equation}
\frac{d g_{jk}}{d t}
= \left[\frac{\partial g_{jk}(x^i,t)}{\partial t} \right]\big|_{x^i \in \Sigma_t}.
\label{g-rate}
\end{equation}

The formula (\ref{holographic-sigma}) relates the entropy production $\sigma^R$ in the field theory and the rate of change of the spacetime metric $g_{jk}(x^i,t)$. In (\ref{g-rate}), $x^i $ ($i=1,\dots d$) denotes the spatial coordinates, and $x^i \in \Sigma_t$ means that these coordinates shall be evaluated on the minimal surface after taking partial time derivative. This equation should be supplemented with the RT equation in order to determine such a surface:
\begin{equation}
\left( \frac{\delta a}{\delta \Sigma_t }\right)_{g_{jk}}=0.
\label{TR2}
\end{equation}
This prescription is useful for the specific example we are investigating. Equation (\ref{holographic-sigma}) can be integrated out in the time parameter to obtain the (induced) metric, during an interval $(t_- , t_+)$ provided that
\begin{equation}
\frac{d\Sigma_t}{dt} = 0,\qquad \forall \; t \in (t_- , t_+) .
\label{trackingcondition}
\end{equation}

In fact, regarding the limit in which $\Omega$ covers the whole sphere $S^d$,
one can do an ansatz for the metric with spherical symmetry and AdS asymptotics. In d+1 dimensions, the spacetime metric is given by
\begin{equation}
ds^2 = -  h(r)dt^2 + \frac{1}{h(r)}dr^2 + g_{\Omega\Omega}(r,t) d\Omega_{d-1}^2,
\label{metrica-h}
\end{equation}
and the spatial metric $g_{jk}$ on ${\cal N}_t $ is nothing but
\begin{equation}
g_{jk} dx^j dx^k \equiv h(r)^{-1}dr^2 + g_{\Omega\Omega}(r,t) d\Omega_{d-1}^2,
\end{equation}
where $h(r)\to R^{-2} r^2$ as $r\to \infty$ is the asymptotic (AdS) condition. In (\ref{metrica-h}), $t$ is the time coordinate, $r$ is the radial coordinate, $\Omega_{d-1}$ stands for the polar coordinates, and $R$ is interpreted as the AdS curvature. The exact AdS spacetime is given by $h= 1 + R^{-2} r^2$ and the induced metric on the spheres $r=\mbox{constant}$ is
$g_{\Omega\Omega}(r,t) d\Omega_{d-1}^2  = r^2 d\Omega^2_{d-1} $. Now assume that $h(r) \neq 0 \;\; \forall \;\; r \geq 0$.\footnote{Here we are using radial coordinates such that $g_{rr}^{-1} = g_{tt} \equiv h(r)$ for simplicity, but this is not crucial and depends on changes of the radial coordinates $r \to r'(r)$.} Therefore, the extremal surface is a sphere $r_0(t)$ whose area is
\begin{equation}
a=\int_{r_0(t)} \sqrt{g^{d-1}_{\Omega\Omega}(r,t)} d\Omega
= \sqrt{g^{d-1}_{\Omega\Omega}(r_0(t),t)} \int_{S^{d-1}} d\Omega_{d-1}
\equiv \sqrt{g^{d-1}_{\Omega\Omega}(r_0(t),t)}\;\;k,
\label{area-esfera}
\end{equation}
with $k \equiv \mbox{Vol}(\Omega_{d-1})$. All the computations made above hold for any spacetime dimension.

If $d=2$, then $k=2\pi$ and equation (\ref{holographic-sigma}) reduces to
\begin{equation}
\dot{s} =
\int_{r_0(t)} \frac{1}{2\sqrt{g_{\Omega\Omega}(r,t)}}
\left[\frac{\partial g_{\Omega\Omega}(r, t)}{\partial t} \right]_{r=r_0(t)}
d\Omega = \frac{k}{4G_N} \frac{1}{2\sqrt{g_{\Omega\Omega}(r_0(t),t)}}
\left[\frac{\partial g_{\Omega\Omega}(r, t)}{\partial t} \right]_{r=r_0(t)} ,
\label{eqformetrics}
\end{equation}
where $r_0$ would be the position of the minimal surface. Let us propose the following ansatz for the solution
\begin{equation}
g_{\Omega\Omega}(r,t)= r^2 + \alpha(t),\qquad \alpha \geq0,
\label{solution1}
\end{equation}
Then, equation (\ref{TR2}) gives
\begin{equation}
r_0 = 0.
\end{equation}
So, condition (\ref{trackingcondition}) is satisfied since $\Sigma_t$ is given by the constant embedding $\Omega \mapsto \left( r(\Omega) =0 ,
\Omega \right) \in {\cal N}_t$. Then equation (\ref{eqformetrics}) becomes
\begin{equation}
\frac{d}{d t}\left[\mathfrak{f}(t)\sinh^2 (\gamma t) + \mathfrak{g}(t)\right] =
\frac{k}{4G_N} \frac{1}{2\sqrt{\alpha(t)}}\dot{\alpha}(t) =
\frac{k}{4G_N} \,\frac{d}{d t}\,\sqrt{\alpha(t)},
\label{eqformetrics-ex1}
\end{equation}
whose solution is
\begin{equation}
\alpha(t) =  \left(\frac{4G_N s(t)}{k}\right)^2.
\label{solution-ex1}
\end{equation}
Thus, in principle, the function $h(r)$ could be determined by solving the
Einstein equations (EE) with this input, and the method would allow to determine
the space metric from the behavior of the extremal surface.

As it will be clear in the next section, the later times behavior of the entropy indicates that  the system thermalizes (see equation(\ref{at}))
\cite{cardy-calabrese, malda-hartman}. So, the later times regime might be holographically modeled by a two sided geometry consisting of slight (time dependent) deformations from the maximally extended AdS-Schwarschild solution, whose boundaries (where $\Phi$ and $\Psi$ live) are causally disconnected by an event horizon (see reference \cite{eternal}), at least classically. The following model is going to be built up more
precisely along these lines of thought.
\begin{figure}[!h]
\centering
\includegraphics{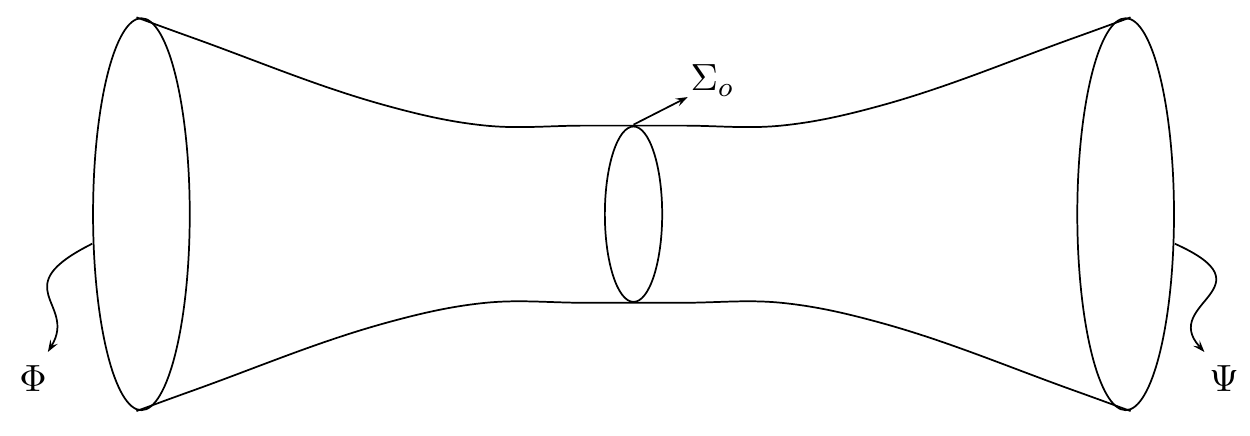}
\caption{Spatial section t = constant of the wormhole solution.}
\label{fig2}
\end{figure}

The example above is a one-sided geometry that can be interpreted as the dual to the state described by the reduced density matrix
$\rho_\Phi \equiv Tr_\Psi |0(t)\rangle \langle 0(t)|$ in the field theory. This
can also be expressed as a two-sided geometry dual to the full state
$|0(t)\rangle \in {\cal H}_\Phi \otimes {\cal H}_\Psi$. For example, for the AdS-Schwarzschild solution with $h(r) = 1- 2M/r^{d-2} + R^{-2} r^2$, this state is the
thermal matrix density $ \rho(\beta)$, while its TFD representation
$|0(\beta)\rangle$ corresponds to the Kruskal extension of the solution (see the construction of \cite{eternal}). This type of spacetimes describes a sort of
wormhole such that the minimal area surface is clearly placed at the throat of the geometry (figure \ref{fig2}).

Moreover, if the metric is globally a General Relativity (GR) solution (for some physical energy-momentum tensor), the topological censorship theorems \cite{tcensorship} state that there is an event horizon separating causally these two conformal boundaries placed at $r \sim \pm\infty$. Nevertheless, if the boundary field theories are assumed to interact as in the example above (equation (\ref{solution-ex1})), then the boundaries should be causally connected (see \cite{botta-arias-silva}).  Classically, this conflict can be bypassed only if we assume that the solution above, with $h(r)$ maximally extended to all the real values of $r$ with a throat of radius $\sqrt{\alpha}$ at $r=0$, does not contain event horizons anywhere; that is, it can be only a solution of some deformation from GR dynamics (e.g., Lovelock theory \cite{lovelock}, Lorentz violating gravities \cite{lorentz-violate}, etc.).  However, a mechanism that makes the black hole traversable, based on a teleportation protocol, was presented in \cite{maldanovo}. This mechanism is suitable to understand the holographic picture for the entropy in the asymptotic limit ($\gamma t \gg 1$).

\section{Constructing a holographic dual model}

Different types of dual geometries corresponding to the behavior described in the previous section could be built up. As  emphasized in \cite{Marika}, it is not
clear whether a system of two interacting CFTs can be realized holographically. In the case studied here, we have two peculiarities: a time-dependent entropy and the breaking of Lorentz symmetry by dissipative processes. In fact, we will present a different scenario of those studied in \cite{TakacoupledCFTs, mozafar} and \cite{Marika}. Let us shed some light on the kind of model we intend to present. From equation (\ref{et}), it can be seen that the early times behavior of the entropy is similar to the results of \cite{TakacoupledCFTs, mozafar}, where it was calculated the entropy for two interacting theories. On the other hand, the asymptotic behavior of the entropy, defined in equation (\ref{at}), suggests that there is thermalization.  So, in the later times, there is a formation of a horizon.

\subsection{Early times}
In early times, the geometry of a spatial slice should be schematically similar to
that shown in figure \ref{fig2}, with two locally asymptotically AdS regions such
that the two fields $\Phi$ and $\Psi$ live on the respective asymptotic
boundaries. Notice that the time dependent state is $|0(\gamma t)\rangle$, which, according to the gauge/gravity correspondence, is dual to a metric that depends on time in the same way, that is, $g_{\mu \nu} = g_{\mu \nu}(\gamma t)$; also, for $\gamma =0$, the field theory is conformal. Since for the time dependent state $\gamma = 0$ is equivalent to $t=0$, the system remains in its fundamental state $|0\rangle\otimes|0\rangle$, which is dual to the geometry described precisely by two disconnected AdS spaces as shown in figure \ref{fig1}\footnote{The holographic interpretation of these disconnected geometries have been discussed in \cite{collapse, VR}.}. We can encompass these two disconnected copies in the same expression for the metric by the  introduction of a ``radial" global coordinate $\varrho$, defined in the intervals $\varrho < 0$ and $\varrho > 0$, respectively\footnote{The coordinate $\varrho$ is defined in terms of the usual global radial coordinate $\chi$ as $\cosh^2 \chi \equiv \displaystyle\frac{1}{1-\varrho^2}$.}. Both manifolds are analytically completed by adding the limit points $\varrho \to 0^\pm$. Expressed  in terms of these coordinates, the metric in $d+1$ dimensions reads
\begin{eqnarray}
ds^2=R^2\left[-\frac{dt^2}{1-\varrho^2}+\frac{d\varrho^2}{(1-\varrho^2)^2}
+\frac{\varrho^2 + \epsilon^2}{1-\varrho^2}d\Omega^2_{d-1}\right],
\label{global-ads}
\end{eqnarray}
where $\Omega^2_{d-1}$ denotes the metric on a compact Euclidean manifold and we are taking the limit $\epsilon \rightarrow 0$. The two asymptotic boundaries correspond to $\varrho =\pm 1$. In the limit $\epsilon \rightarrow 0$, this is the dual geometry for $t=0$ describing two AdS spaces in global coordinates, and there is no causal curves connecting both manifolds. We argue that when  the interaction  is turned on (or $t\neq 0$), there is formation of a throat that entangles the vacuum  dynamically.  In this scenario we have a topology change ($\epsilon \neq 0$ regime), and the two separate spaces are now connected. This mechanism is hard to understand  using a  GR  solution;  we need some GR deformation or some quantum gravity effect that are beyond the scope of this work. We are not going to draw a holographic picture of the early times entropy and we will concentrate just in the later times.

\subsection{Later times}

The later times' entropy is ruled by equation (\ref{at}). Note that, if $\gamma$ is proportional to the temperature, we get the Cardi-Calabrese's result, suggesting thermalization.  This implies that the holographic dual of this theory is BTZ black hole and the fields live, respectively, on the two asymptotically AdS boundaries. As shown in \cite{TraversableWormholes} and \cite{maldanovo},  it is possible to make the wormhole traversable by attaching a coupling between the fields. Effectively, somehow there is an attractive force between the boundaries that makes the radius of the event horizon to decrease,  exposing the interior of the black hole. The process can be viewed as a teleportation protocol that sets up a quantum coupling of the form $e^{igO_AO_B}$, where $O_A$ and $O_B$ represent operators of the two boundary theories.  Let us show that we have a similar structure.  The (leading first-order in gamma) interaction term between the boundaries is
\begin{equation}
H^1_{int} = i\frac{\gamma}{2}\;\sum_{k} \;  \left(A_{-k} B_{k}
- A^{\dagger}_{-k} B^{\dagger}_{k}\right).
\label{hintp}
\end{equation}

In fact, the theory studied here can be seen as a deformation in two decoupled conformal field theories living on the two boundaries, namely
\begin{equation}
\left\langle e^{\,\,g\int_{\partial M}O_A O_B} \right\rangle_{\beta} = Z_{CFT} (g,\beta),
\end{equation}
and the mechanism of information exchange proposed in \cite{TraversableWormholes}, and studied in \cite{maldanovo}, works for $g>0$. The indice $\beta$ in this equation means that the expectation value is taken on the TFD vacuum.  By using the BDHM (Banks, Douglas, Horowitz, Martinec) recipe \cite{BDHM, BDHM2}, one can quantize a field near the boundary of the AdS black hole, and $O_A$, $O_B$ can be some linear combination of $A^{\dagger}_{-k}$, $A_k$ and $B^{\dagger}_{k}$, $B_{-k}$ respectively.  In particular, if we define hermitian operators as
\begin{eqnarray}
O_A &=& \frac{A^{\dagger} - A}{2i}, \qquad O'_A = \frac{A+ A^{\dagger}}{2},
\nonumber
\\
O_B &=& \frac{B^{\dagger} - B}{2i},  \qquad O'_B = \frac{B+ B^{\dagger}}{2}.
\end{eqnarray}
the hamiltonian (\ref{hintp}) has the form $H= \frac{\gamma}{2}(O_AO_B + O'_AO'_B)$, according to prescription of \cite{maldanovo}. Finally, we have
\begin{eqnarray}
\left\langle 0 (\beta) \right| \exp(-i H^1_{int} \: t) \left|0 (\beta) \right\rangle
 = \left\langle 0 (\beta) \right|\exp\left[\sum \; \frac{\gamma t}{2} \left( O_A O_B
+ O'_AO'_B \right)\right]
\left|0 (\beta) \right\rangle ,
\end{eqnarray}
and then $g \equiv \frac{\gamma}{2}$ is positive such as we have thought.
	
Based on these discussions, we will present in this section a gravitational system that fits to our dissipative theory. In particular, it has similar field's equation of motion and it is described in a coordinate system covering a region inside the event horizon.

\subsection{The BTZ black hole with a time-dependent boundary}

Although the results presented so far do not depend on the spacetime dimension, we will now show that in two dimensions it is possible to obtain a clearer geometric interpretation of the entanglement entropy we have calculated. It is well known that, at finite temperature, two dimensional CFT is dual to a BTZ black hole. In general, starting from a bulk metric with a flat boundary, it is possible to perform a coordinate transformation that generates a conformally flat boundary. In order to have a simple way to use the holographic renormalization and to calculate the stress-energy tensor \cite{Skenderis, sken1}, the bulk metric is written in Fefferman-Graham coordinates \cite{FG}. In reference \cite{Tetradis}, this procedure was carried out using a particular coordinate transformation, in order to produce boundary metrics with conformal  factors that have explicit time dependence, such that the  boundary metric is of the Friedmann-Robertson-Walker (FRW) type
\begin{equation}
ds^2= (-dt^2 + a^2(t)d\Omega_D^2).
\end{equation}

For $AdS_5$ black holes, $D=3$, it is possible to perform a time transformation such that this  system perfectly fits the dissipative system presented here. Using the conformal time $\eta= \displaystyle\int \displaystyle\frac{1}{a(t)}$, the D-dimensional boundary metric has the conformal form $ds^2= a(\eta)^2(-d\eta^2 + d\Omega_D^2)$. The equation of motion becomes
\begin{equation}
(\partial_{\eta}^2- \nabla^2)\Phi +(D-2)\frac{\dot a(\eta)}{a(\eta)} \dot{\Phi} = 0,
\end{equation}
where the dot is the derivative relative to $\eta$. Now, this  equation of motion is exactly the dissipative equation of motion, where for $CFT_4$  we have, for $a(\eta)= e^{H \eta}$, $H= \frac{\gamma}{2}$.   For $AdS_3$ an approximation must be done. We are going to focus on the  $AdS_3$ case, where the coordinate transformation is produced in an analytical form and it is easier to get a minimal area in order to use the RT formula.

We will discuss in this section the main results of \cite{Tetradis}. We start with the BTZ metric
\begin{equation}
ds^2 = -f(r) dt^2 + \frac{dr^2}{f(r)} + r^2 d\phi^2 \:, \:\:\: f(r) = r^2 - \mu , \label{btz}
\end{equation}
where the temperature and the entropy are given by
\begin{equation}
T = \frac{1}{2\pi} \sqrt{\mu} \:, \:\:\: S = \frac{V}{4G_3} \sqrt{\mu} .
\end{equation}
As mentioned earlier, for our holographic applications it is easier to define Fefferman-Graham type coordinates. Defining the variable z
\begin{equation}
z= \frac{2}{\mu}(r-\sqrt{r-\mu}) ,
\end{equation}
the metric takes the form
\begin{equation}
ds^2 = \frac{1}{z^2} \left[dz^2 - \left(1-\frac{\mu}{4} z^2 \right)^2 dt^2 + \left(1+\frac{\mu}{4} z^2 \right)^2 d\phi^2\right] .
\label{mfg}
\end{equation}
In reference \cite{Tetradis}, it is shown that it is possible to write this metric as follows
\begin{equation}
ds^2 = \frac{1}{z^2} [dz^2 - \mathcal{N}^2 (\tau, z) d\tau^2 + \mathcal{A}^2 (\tau, z) d\phi^2 ] ,
\label{mt}
\end{equation}
with
\begin{eqnarray}
\mathcal{A} (\tau, z) = a (\tau) \left(1 + \frac{\mu-\dot{a}^2 (\tau)}{4a(\tau)^2} z^2 \right)\\
\mathcal{N} (\tau, z) = 1 - \frac{\mu - \dot{a}^2 + 2 a \ddot{a}}{4a^2} z^2 = \frac{\dot{\mathcal{A}} (\tau, z)}{\dot{a}} ,
\label{mta}
\end{eqnarray}
for some function $a(t)$ constrained by Einstein equations (see \cite{Tetradis}  for details).  Comparing  (\ref{mt}) with (\ref{mta}), we have
\begin{equation}
r(\tau, z) = \frac{\mathcal{A} (\tau, z)}{z} = \frac{a}{z} + \frac{\mu - \dot{a}^2 z}{4} \frac{z}{a} .
\end{equation}
Now, taking the limit $z \rightarrow \infty$,  we have a time-dependent boundary metric of the FRW type
\begin{equation}
ds^2= d\tau^2- a(t)d\phi^2 ,
\end{equation}
where $\phi$ is a periodic variable. By coupling a scalar field to this metric and neglecting the self-interaction of the field, we have the equation of motion
\begin{equation}
\left(\partial_t^2- \frac{1}{a^2}\nabla^2 \right)\phi+ H\partial_t\phi =0 ,
\label{eqfrw}
\end{equation}
where $H = \displaystyle\frac{\dot a}{a}$.

Note that, in general, the term $H\dot{\phi}$ is absorbed in the frequency term by using the conformal time variable and rescaling the fields. However, if we keep the structure of equation (\ref{eqfrw}), the approach of doubling the degrees of freedom to make canonical quantization used in \cite{vitiello} allows to explore the subtleties involving the definition of the vacuum of the dissipative theory. In our case, the auxiliary system is in fact a physical system, defined as the degrees of freedom behind the horizon.

Before discussing the kind of approximation we intend to do in order to use this system to give a geometric interpretation of the previous one, let us see another important feature of the coordinate system defined in \cite{Tetradis}. As in the static case, the coordinates $(\tau, z)$ do not span  the full BTZ geometry. The horizon is defined by $r_e= \sqrt{\mu}$.  If we find the two roots of the equation  $r(z,\tau)= r_e$, we discover that the coordinate system
covers the two regions outside the event horizons, located at
\begin{eqnarray}
z_{e1} = \frac{2a}{\sqrt{\mu} + \dot{a}} ,
\\
z_{e2} = \frac{2a}{\sqrt{\mu} - \dot{a}} .
\end{eqnarray}
This is a very important result, because this implies that, for constant $\tau$, the minimal value of $r(\tau,z)$ is
\begin{equation}
r_m= \sqrt{\mu -\dot{a}^{2}} .
\label{rm}
\end{equation}
Therefore, the lowest value of $r$ is less than $r_e$ and  the boundary system ``sees'' inside the horizon.
According to RT prescription, in this coordinate system we have the following equation for the entanglement entropy \cite{Tetradis}:
\begin{equation}
S = \frac{1}{4G_3} r_m .
\end{equation}
Note that this result is in agreement with the reference \cite{malda-hartman} where, for non static situations, the minimum area is behind the event horizon.

\section{Gravity computations}

In this part, we explain an approximated holographic method to analyze the DCFT. It assumes a holographic model for our dissipative system by a little time interval around (before) certain specific instant of its evolution $t_0$, belonging to certain (stable) regime of interest.

Let us use a BTZ black hole in the  coordinates discussed in the previous section, simply referred as $a$-coordinates. For this calculation, let us assume that there exists a $t_0 \in \Re$ such that the holographic model simply stops to be a suitable description of our system, so the physical time parameter is taken to be $t \leq t_0$.  We are going to assume that $\mu$ in (\ref{btz}) varies slowly over time, such that the BTZ is in fact a sort of Vaidya solution in the adiabatic approximation, where $\displaystyle\frac{\dot{\mu}}{\mu} \approx 0$ \cite{Ziogas,covariantEntropy} (see appendix). The Vaidya solution is the simplest example of a time dependent gravitational background which is explored in the context of the RT conjecture \cite{covariantEntropy, timedependent, Hubeny:2006yu}. It represents the black hole collapse and has a time dependent horizon defining a time dependent temperature \cite{SQW,ECVD,Li,XLi}.   We assume also that for the $AdS_3$, the holographic model describes only the system approximately, i.e. in a neighborhood (immediately before) of $t_0$, so to first order
\begin{eqnarray}
\mu = \mu_0 + (t-t_0) \mu_1 + o^2 ,
\label{muBH}\\
a = a_0 + (t-t_0) a_1 + o^2 .
\label{aBH}
\end{eqnarray}
In order to describe our dissipative system as a scalar field propagating on anti-de Sitter boundary, we must demand also that
\begin{equation}
\dot{a}/ a  \approx \gamma  \,\,\,, \,\, a\approx 1, \,\,\,\, \forall t\approx t_0
\end{equation}
so the equation of motion (\ref{eqfrw}) agrees with that of our original system.

In this gravity model, entanglement entropy can be computed using the Ryu-Takayanagi formula. As the fields are defined in all  space spanned by the board coordinate system, the minimal area to use in the Ryu-Takayanagi formula is the area computed with  the minimal value of $r(\tau,z)$, written in (\ref{rm}). The entropy is
\begin{equation}
s=\frac{1}{4G_3}\sqrt{ \mu - \dot{a}^2} .
\label{entanglementS}
\end{equation}
For a $t_0 \gg \gamma ^{-1}$, this shall be demanded to agree with $s \sim \gamma t$ (therefore, the first order expansion of (\ref{entanglementS})), and using (\ref{muBH}) and (\ref{aBH}) we obtain
\begin{eqnarray}
\mu_1 = 8G_3 \gamma \sqrt {\mu_0 - \gamma^2} ,
\label{mu1} \\
t_0 =  \frac{1}{4G_3} \gamma^{-1} \sqrt {\mu_0 - \gamma^2} ,
\label{t0}
\end{eqnarray}
which implies the physical bound
\begin{equation}
\mu_0\geq\gamma^2 .
\label{bound}
\end{equation}
Our description would be inconsistent otherwise. Now, the limit $\gamma t_0 >> 1$ corresponds to small $G_3$, implying the large N limit, since, according to the  AdS/CFT dictionary, $G_3 \propto 1/N^2$. These results mean that the spacetime parameters are determined for the parameters of our model up to first order (in a Taylor expansion on time). Note that the adiabatic approximation corresponds to the constraint  $\mu_1/\mu_0 \sim \gamma G_3 \ll 1$, again according to large N limit. All the approximations are consistent with  small $G_3$ limit.

\subsection{On the holographic energy density }

Using the typical AdS/CFT dictionary equation
\begin{equation}
\langle T^{(CFT)}_{\mu\nu} \rangle = \frac{1}{8\pi G_3} [g^{(2)} - tr (g^{(2)}) g^{(0)}] ,
\end{equation}
where
\begin{equation}
g_{\mu\nu} = g^{(0)}_{\mu\nu} + z^2 g^{(2)}_{\mu\nu} + z^4 g^{(4)}_{\mu\nu} ,
\end{equation}
we have the following equation \cite{Tetradis}
\begin{equation}
\rho = \frac{E}{V} = - \langle T^{\tau}_{\:\:\:\tau}\rangle = \frac{1}{16\pi G_3}\left( \frac{\mu - \dot{a}^2}{a^2}\right) ,
\label{eqph}
\end{equation}
where $ \rho $ is the energy density for the boundary theory. Using the linear approximation, it follows
\begin{equation}
\rho = \frac{1}{16\pi G_3} (\frac{\mu - \dot{a}^2}{a^2} )\approx \frac{1}{16\pi G_3}  (\mu_0 - \gamma^2 \,\, + o(t- t_0)) ,
\end{equation}
so that our bound (\ref{bound}) is consistent with the non-negativity of this quantity. Compatible with dissipative behavior, the energy of the actual system decreases as
\begin{equation}
\frac{d\rho}{dt} =
\frac{1}{16\pi G_3}  ( \mu_1 - 2\gamma (\mu_0 - \gamma^2)) + o(t- t_0) ,
\end{equation}
near the final time $t_0$. Then, using (\ref{mu1}),
\begin{equation}
\dot{\rho}_0  = \frac{d \rho}{dt} (t = 0) \approx 2\gamma \frac{1}{16\pi G_3} \left(4G_3 \sqrt{(\mu_0 - \gamma^2)}-(\mu_0 - \gamma^2)\right) ,
\end{equation}
which is negative (decreasing) as $(\mu_0 - \gamma^2) >0$ for smal $G_3$.  Therefore, let us assume that it can decay up to some minimal value, say  $\rho_0$,
\begin{equation}
\rho_0 = \frac{1}{16\pi G_3} (\mu_0 - \gamma^2)\geq 0 .
\end{equation}

\subsection{Relating $\gamma$ with  temperature}

Now we are going to show how the coupling $\gamma$ can be holographically interpretated as the temperature. In the adiabatic approximation, as the BTZ black hole radius slowly varies over time, the  black hole stays in equilibrium during the whole evolution and it  is possible to define a time dependent temperature. For the Vaidya/BTZ black hole, the temperature is
\begin{equation}
T = \frac{1}{2\pi} \sqrt{\mu} \approx \frac{1}{2\pi} \left(\sqrt{\mu_0} + \frac{\mu_1}{2 \sqrt{\mu_0}} (t- t_0) + o^2 \right) .
\end{equation}
Although it grows with time, its approximated value as $t\to t_0$ is
\begin{equation}
T_0 (\rho_0) =\frac{1}{2\pi} \sqrt{\mu_0 (\rho_0)} .
\end{equation}
Notice that we have expressed this as a function of the extremal CFT energy density  achieved earlier. Let us observe that, in the large N limit, we get a suggestive result:
for $\gamma \gg   G_{3}\rho_{0}$,
\begin{equation}
T_0  = \frac{1}{2\pi} \sqrt{\gamma^2 + 16 \pi G_3 \rho_0}
\approx \frac{\gamma}{2\pi} + 4 G_3\frac{\rho_0}{\gamma} ,
\end{equation}
which exactly coincides with the Cardy-Calabrese result.

The last term denotes the subleading contribution, which is proportional to
$\frac{\rho_0}{\gamma}$.
This calculation can be viewed as a holographic check on the consistency that the entanglement in DCFT behaves as the entanglement of an ordinary CFT at finite temperature \cite{cardy-calabrese}.

\subsection{The approximated state, its time evolution and the breakdown of our $AdS_3/DCFT_2$ description}

Our approach discussed in the last subsection can be described as an approximation on the entangled state.

We know that the BTZ spacetime corresponds to the TFD thermal state $\left|0_\beta \right\rangle $ in the boundary theory \cite{eternal}. Thus, the state corresponding to the gravitational system should be represented as 
\begin{equation}
\left|0_{a(t)} \right\rangle =
G_{a(t)} \left|0_\beta \right\rangle = G_{a(t)} G_{\beta} |0\rangle ,
\label{proposalT}
\end{equation}
where $G_{a(t)}$ represents a (unknown) unitary transformation associated to certain specific conformal transformation in CFT,
such that in the limit $a(t)\to 1$, $G_{a(t)}\to \mathbb{1}$. $G_\beta$ is the usual thermal Bogoliubov transformation. Actually, we propose that our DCFT system, described by equation (\ref{res1}), is approximately given by the state (\ref{proposalT}),
\begin{equation}
\left|\Psi(t) \right\rangle \approx \left|0_{a(t)} \right\rangle
= G_{a(t)} \left|0_\beta \right\rangle
= G_{a(t)}  G_{\beta} \left|0 \right\rangle ,
\label{proposalT-approx}
\end{equation}
as $t$ approaches $t_0$. This (approximated) equation allows us to interpret correctly our previous calculations, and to understand correctly the ranks of validity. Notice that the state on the right-hand side $\left|0_{a(t)}\right\rangle$ is the conformal ground state dual to the BTZ, in $a(t)$-coordinates.

As said before, $G_{a(t)}$ represents a (unknown) unitary transformation associated to a specific conformal transformation in CFT, such that in the limit $a(t)\to 1$, $G_{a(t)}\to \mathbb{1}$; $G_\beta$ is the usual thermal Bogoliubov transformation. Therefore,
\begin{equation}
\left|\Psi(t) \right\rangle = U(t - t_0) \left|\Psi (t_0)\right\rangle .
\end{equation}
Then, using (\ref{estadoentropia}), the time evolution can be expressed as generated by the entropy operator as
\begin{equation}
\left|\Psi(t) \right\rangle
= \exp{\frac{1}{2}[S(t_0) - S(t)]}\left|\Psi (t_0)\right\rangle .
\label{evolutionS}
\end{equation}
This equation can be checked in the geometric dual through the RT formula, i.e,
\begin{equation}
\mbox{Tr} [\rho(t) - \rho(t_0)] S(t_0) = \frac{1}{4G} [a_{min}(t) - a_{min}(t_0)] ,
\end{equation}
where $\rho$ is the reduced density operator computed by tracing out the B's degrees of freedom. The idea is that the left-hand side is computed explicitly for the state $\Psi$, while the right-hand side is computed for $\left|0_{a(t)}\right\rangle$, whose entanglement entropy can be computed directly from the formula (\ref{entanglementS}).

The remark here is that, for parameters given explicitly by equations (\ref{mu1}) and (\ref{t0}), both sides match perfectly. This supports that the approximation (\ref{proposalT-approx}) is, in a sense, consistent with the time evolution of the states. However, the dual gravitational description must break down for very large time. From the point of view of the gravitational solution, if one considers a BTZ solution with very high temperature $T \gg  \sqrt{\mu_0}$, then
\begin{equation}
\mu_0 \gg \gamma^2 ,
\label{thermalphase}
\end{equation}
and the formula for the entanglement entropy gives $s= \displaystyle\frac{2\pi}{4G_3} \sqrt{\mu_0}$, which coincides with the thermodynamic entropy, and therefore $s \propto T $. So, the state corresponding to this solution is nothing but the TFD thermal vacuum \cite{eternal}:
\begin{equation}
\left|0(\beta) \right\rangle
= Z^{-1}(\beta)\prod_{k}
\exp\left[e^{-E_k \beta/2}
A^{\dagger}_{-k}B^{\dagger}_{k}\right]\left| 0 \right\rangle ,
\label{TFDstate}
\end{equation}
while the state of our DCFT system (\ref{res1}) is manifestly different, showing that for very large black hole temperature, our original approximation (\ref{proposalT-approx}) breaks down.

Nevertheless, the form of the states (\ref{res1}) and (\ref{TFDstate}), and the configuration of the systems are suggestively similar in many aspects. So a natural question that arises here is if there exists some way of slightly modifying the action/states of the double CFT proposed, in order to capture more of the dual gravitational description. The simplest possibility is to introduce a parameter, interpreted as a temperature from the A-subsystem, such that: (a) in the limit $\beta \to 0$, the state agrees with \ref{TFDstate},  and (b) for low (temperature) scales $\beta^{-1} \to \,\sim \gamma$, one would recover the (main) dissipative/damping  behavior (\ref{res1}). In the limit $\gamma t \to 0$, the (sub)systems $A$ and $B$ decouple, and they can be seen as the standard  TFD duplication in the state (\ref{TFDstate}). This will be studied in a forthcoming work.

\begin{figure}[!h]
\centering
\includegraphics{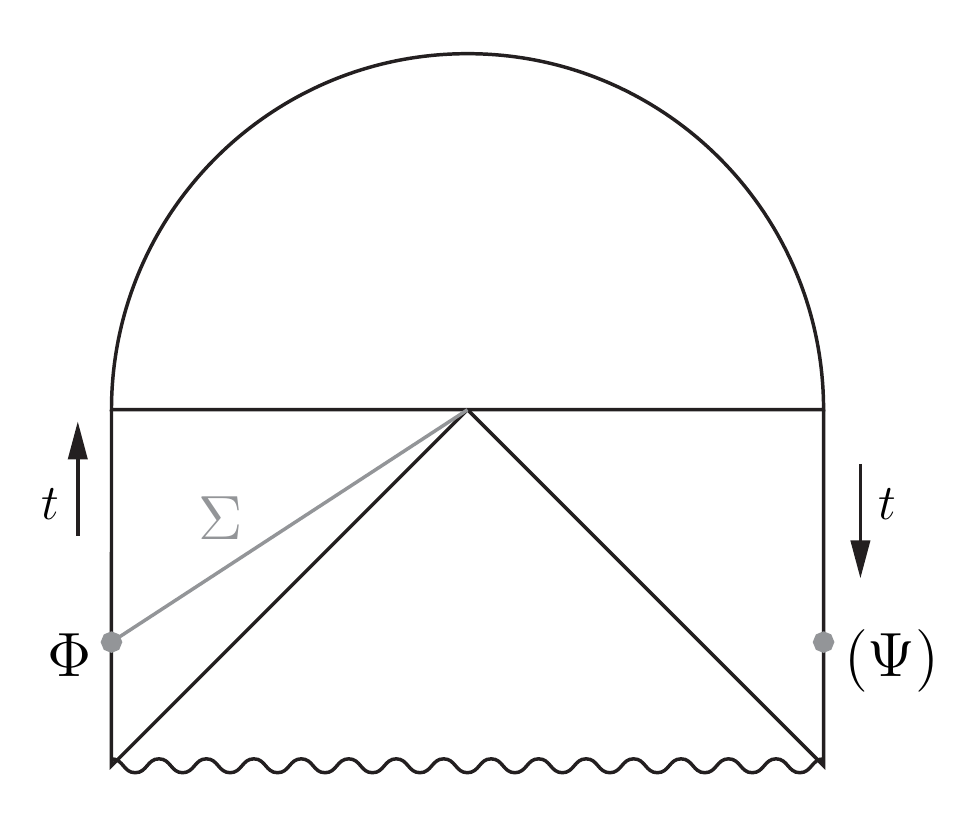}
\caption{The figure represents schematically the geometric dual used to (approximately) describe the system. The upper part represents the black hole state (see \cite{eternal}) described by (a half of) the Euclidean BTZ solution in the Hartle-Hawking wave functional. The horizontal line corresponds to $t_0$. Our system is approximately described by the solution on an evolving surface $\Sigma(t\leq t_0)$.}
\label{fig5}
\end{figure}

\section{Conclusion}

In this work, we have studied a close relation  between dissipation and entanglement in the context of AdS/CFT correspondence. This kind of relation was experimentally verified in \cite{EntanglementDissipation}, where it was shown that dissipation generates continuously entanglement between two macroscopic objects. In order to study quantum dissipation in conformal field theories, we have followed the strategy of \cite{vitiello, parentani, adamek}, which consists of doubling the degrees of freedom of the original system by defining an auxiliary one. In particular, using the canonical formalism presented in \cite{vitiello}, we have defined an entropy operator, responsible of controlling the dissipative dynamics. The entropy operator, which naturally appears in the scenario  studied here, turns up to be in fact  the modular Hamiltonian. A geometrical and physical interpretation of the auxiliary system and entropy operator is given in the context of the AdS/CFT correspondence using the RT formula. One has two asymptotically AdS regions such that the two theories live on the respective asymptotic boundaries. The dissipation at one boundary is interpreted as an exchange of energy with the other boundary, which is controlled by a kind of constant coupling. The scenario is similar to the one presented in \cite{collapse}, where a  teleportation protocol is explored in order to make the wormholes traversable,  introducing interaction beween the two asymptotic boundaries.

We have showed that the vacuum state evolves in time as an entangled state and the entanglement entropy was calculated. The entropy depends on time and has two distinct behaviors: in the early times, it is the typical entanglement entropy of two interacting theories; in the later times, the entropy's behavior suggests thermalization. In order to give a geometric interpretation, we have looked for a gravitational system that reproduces a similar equation of motion in the boundary. Since this kind of dissipative behavior was studied long time ago in the context of  scalar fields coupled to Friedmann-Robertson-Walker (FRW) metric with a damping term coming from the Hubble constant \cite{dissFRW, grazi1, grazi2, buryak, habib}, then the BTZ black hole in the coordinate system developed in \cite{Tetradis} is the natural system to study here, as it has a FRW metric in the boundary. However,  in two dimensions a linear aproximation must be done. In addiction, owing to the linear time dependence of the entropy, we have showed that the dual geometry must be not the BTZ black hole in the coordinates defined in \cite{Tetradis}, but  an adiabatic approximation for the Vaidya black hole defined in the same coordinate system.

We have used the RT formula to relate the BTZ/Vaidya geometry's throat area to the entropy.  As the time evolution of the entangled state is controlled by the entropy operator, the RT conjecture allows us to relate the time evolution of the throat with the time evolution of the vacuum state.  It should be emphasized that this kind of relation involving state/area is possible because, in the canonical formalism used here, there is a well defined  entanglement entropy operator. A possible close relation of this operator with a kind of  area operator, defined in the context of quantum gravity, deserves further development (for some suggestions in this sense, see for instance \cite{area-qg}).

It was shown that the approximation is consistent with the time evolution of the states. However, the dual gravitational description must break down for very large time, when the entanglement entropy calculated here is in fact a thermodinamical entropy of  BTZ solution with very high temperature $T \gg  \sqrt{\mu_0}$. In this limit, the asymptotically entangled state (\ref{res1}) must be the TFD thermal vacuum (see figure \ref{fig5}). This structure suggests strongly that one can adopt reciprocal point of view and take seriously the dual gravitational description to reconstruct the dissipative conformal field theory (DCFT); in other words, to find some slightly deformation of the DCFT theory in according to the dual gravitational description. 

In addition to the previously discussed deformation of the DCFT, there are some applications of this work that we intend to do. Another important sequence to this work is to use the approach developed here in the context of Janus solutions of supergravity (studied in \cite{Janus}), where two CFTs defined on $1+1$ dimensional half spaces are glued together over a $0 + 1$ dimensional interface. If we can generalize the Janus deformation in order to include the dissipative deformation studied here, it will be  possible to have a holographic interpretation of the dissipative system in terms of the time dependent Janus' black hole. Also, it will be important to compute the retarded
Green's function in order to compare with the results of reference \cite{Banerjee}.
Finally, in the spirit of reference \cite{maldanovo}, it will be very interesting to study the specific teleportation protocol that reproduces the scenario presented here, as well as to apply the technology developed here in the $AdS_2$ case, in order to find a possible connection between the DCFT and the Sachdev-Ye-Kitaev quantum mechanical model.

\begin{acknowledgments}
M. Botta Cantcheff would like to thank CONICET for financial support. The authors would like to thank Pedro Martinez for the help with the figures.
\end{acknowledgments}

\appendix
\section{Maximally extended geometries and minimal surfaces}

\begin{figure}[!h]
\centering
\includegraphics{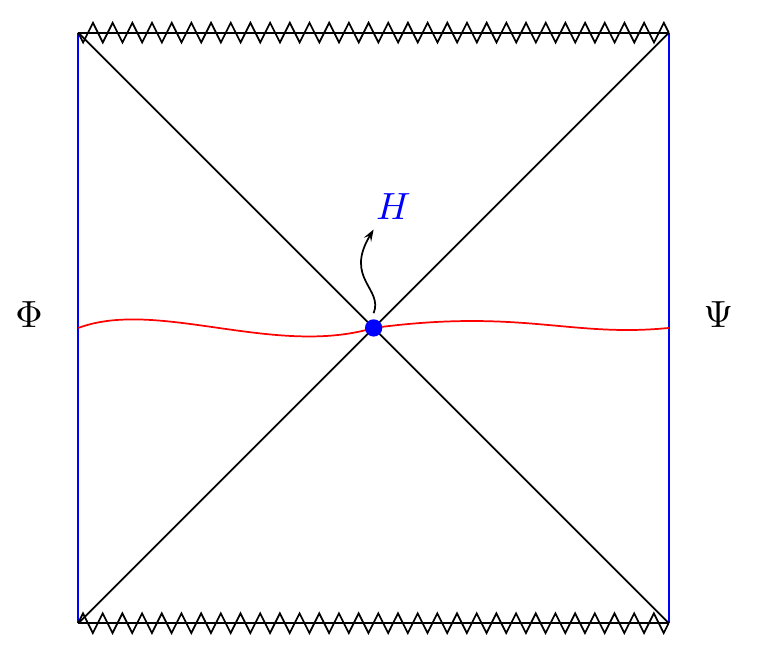}
\caption{The red line represents the spatial slice shown in Figure \ref{fig2}. The point H (in blue) represents the event horizon.}
\label{fig4}
\end{figure}

Let us verify that this type of spacetimes can be described as two-sided geometries through a maximal (Kruskal-like) extension; then, the resulting geometry is a sort of wormhole such that the minimal area surface is clearly seen as the throat of the geometry; see figure \ref{fig4}. This wormhole solution is the analogous to the Einstein-Rosen bridge in four dimensions and it is not traversable classically.

So, let us consider for instance the BTZ solution
\begin{equation}
ds^2 = -  h(r)dt^2 + \frac{1}{h(r)}dr^2 + r^2 d\phi^2 ,
\label{btz-r}
\end{equation}
where $h(r)= r^2 - \mu$. Doing the following coordinate's change
\begin{equation}
\rho^2 \equiv r^2 - \mu ,
\label{coord-change}
\end{equation}
for $r^2\geq \mu$; thus $ \rho d \rho = r dr $ and (\ref{btz-r}) becomes
\begin{equation}
ds^2 = -  \rho^2 dt^2 + \frac{1}{\rho^2 + \mu}d\rho^2 + (\rho^2 + \mu) d\phi^2 .
\label{btz-rho}
\end{equation}
So, the solution (\ref{btz-r}), valid for $ 0 \leq \rho \leq \infty$, can be extended here to all the real line $-\infty \leq \rho \leq \infty$, which describes two causally disconnected asymptotically AdS spacelike regions joined by the surface (throat) $\rho =0$, as shown schematically in figure \ref{fig1}. In addition, notice that, in these coordinates, there is a horizon precisely in $\rho =0$.

By symmetry, the minimal surface is a sphere of codimension $2$, whose radius is given by $(\rho^2 +\mu)$, as it can be seen clearly from expression (\ref{btz-rho}). Therefore, the result is that the minimal sphere is at the horizon $\rho=0$ with radius $\sqrt{\mu}$ and area $a= 2\pi \sqrt{\mu}$.

\vspace{1cm}

\noindent {\bf The time dependent case}

\vspace{.5cm}

A time dependence of the spacetime, with similar causal structure, can be introduced by doing $\mu\equiv\mu(t)$ in this solution, which can be obtained by a coordinate's change from the Vaidya's solution, as long as the adiabatic approximation is valid, i.e.,  whether  $\dot{\mu}(t) / \mu(t)$ is negligible. In particular, the Vaidya's solutions considered in the present paper can be included in this analysis and one obtains the time dependent wormhole metric
\begin{equation}
ds^2 = -  \rho^2 dt^2 + \frac{1}{\rho^2 + \mu (t)}d\rho^2 + (\rho^2 +\mu (t)) d\phi^2 ,
\label{btz-rho-m(t)}
\end{equation}
where, again, the extremal surface corresponds manifestly to $\rho=0$, with radius $\sqrt{\mu(t)}$. Therefore, by taking $\alpha \equiv \mu$, this metric fits into the examples studied in section IV.

Now the extremal curve shall enclose the horizon such that the surface of extremal area becomes the event horizon, whose area is $a_{eh}=2\pi\sqrt{\mu}$. Using the RT formula, we finally have the relation
\begin{equation}
S(t)= \frac{\pi}{2G}\sqrt{\mu(t)} ,
\end{equation}
and once again one gets consistency of the RT formula with the Bekenstein-Hawking law.

Therefore, as shown above, this solution (for slowly varying $\mu(t)$) can be extended to a similar form of the Kruskal one, and the Penrose's diagram corresponds to the maximally extended AdS black hole (figure \ref{fig4}).

\end{document}